\documentstyle[12pt]{article}

\def\d{\partial}
\def\2{\frac{1}{2}}

\def\h{\hat}

\def\l{\lambda}
\def\m{\mu}
\def\n{\nu}
\def\dd{\delta}
\def\ep{\epsilon}

\def\t{\theta}

\def\s{\star}

\def\o{\omega}
\def\O{\Omega}

\def\A{{\mathcal A}}

\def\D{{\mathcal D}}
\def\OO{{\mathcal O}}
\def\Tr{\mathop{\rm Tr}\nolimits}
\def\mod{\mathop{\rm mod}\nolimits}

\def\const{\mathop{\rm const}\nolimits}

\def\beq{\begin{equation}}
\def\eeq{\end{equation}}
\def\ba{\beq\new\begin{array}{c}}
\def\bal{\begin{array}{l}}
\def\eal{\end{array}}
\def\bac{\begin{array}{c}}
\def\eac{\end{array}}

\def\ea{\end{array}\eeq\\}
\def\be{\ba}
\def\ee{\ea}

\makeatletter
\newdimen\normalarrayskip              
\newdimen\minarrayskip                 
\normalarrayskip\baselineskip
\minarrayskip\jot
\newif\ifold             \oldtrue            \def\new{\oldfalse}
\def\arraymode{\ifold\relax\else\displaystyle\fi} 
\def\eqnumphantom{\phantom{(\theequation)}}     
\def\@arrayskip{\ifold\baselineskip\z@\lineskip\z@
     \else
     \baselineskip\minarrayskip\lineskip2\minarrayskip\fi}
\def\@arrayclassz{\ifcase \@lastchclass \@acolampacol \or
\@ampacol \or \or \or \@addamp \or
   \@acolampacol \or \@firstampfalse \@acol \fi
\edef\@preamble{\@preamble
  \ifcase \@chnum
     \hfil$\relax\arraymode\@sharp$\hfil
     \or $\relax\arraymode\@sharp$\hfil
     \or \hfil$\relax\arraymode\@sharp$\fi}}
\def\@array[#1]#2{\setbox\@arstrutbox=\hbox{\vrule
     height\arraystretch \ht\strutbox
     depth\arraystretch \dp\strutbox
     width\z@}\@mkpream{#2}\edef\@preamble{\halign
\noexpand\@halignto
\bgroup \tabskip\z@ \@arstrut \@preamble \tabskip\z@ \cr}%
\let\@startpbox\@@startpbox \let\@endpbox\@@endpbox
  \if #1t\vtop \else \if#1b\vbox \else \vcenter \fi\fi
  \bgroup \let\par\relax
  \let\@sharp##\let\protect\relax
  \@arrayskip\@preamble}
\def\eqnarray{\stepcounter{equation}%
              \let\@currentlabel=\theequation
              \global\@eqnswtrue
              \global\@eqcnt\z@
              \tabskip\@centering
              \let\\=\@eqncr
              $$%
 \halign to \displaywidth\bgroup
    \eqnumphantom\@eqnsel\hskip\@centering
    $\displaystyle \tabskip\z@ {##}$%
    \global\@eqcnt\@ne \hskip 2\arraycolsep
         $\displaystyle\arraymode{##}$\hfil
    \global\@eqcnt\tw@ \hskip 2\arraycolsep
         $\displaystyle\tabskip\z@{##}$\hfil
         \tabskip\@centering
    &{##}\tabskip\z@\cr}
\begingroup\ifx\undefined\newsymbol \else\def\input#1 {\endgroup}\fi
\newfont{\hr}{msbm10}
\newfont{\ams}{msam10}

\font\numbers=cmss12
\font\upright=cmu10 scaled\magstep1
\def\stroke{\vrule height8pt width0.4pt depth-0.1pt}
\def\topfleck{\vrule height8pt width0.5pt depth-5.9pt}
\def\botfleck{\vrule height2pt width0.5pt depth0.1pt}
\def\Zmath{\vcenter{\hbox{\numbers\rlap{\rlap{Z}\kern 0.8pt\topfleck}\kern
2.2pt
                   \rlap Z\kern 6pt\botfleck\kern 1pt}}}
\def\Qmath{\vcenter{\hbox{\upright\rlap{\rlap{Q}\kern
                   3.8pt\stroke}\phantom{Q}}}}
\def\Nmath{\vcenter{\hbox{\upright\rlap{I}\kern 1.7pt N}}}
\def\Cmath{\vcenter{\hbox{\upright\rlap{\rlap{C}\kern
                   3.8pt\stroke}\phantom{C}}}}
\def\Rmath{\vcenter{\hbox{\upright\rlap{I}\kern 1.7pt R}}}
\def\Z{\ifmmode\Zmath\else$\Zmath$\fi}
\def\Q{\ifmmode\Qmath\else$\Qmath$\fi}
\def\N{\ifmmode\Nmath\else$\Nmath$\fi}
\def\C{\ifmmode\Cmath\else$\Cmath$\fi}
\def\R{\ifmmode\Rmath\else$\Rmath$\fi}
\def\numberbysection{\@addtoreset{equation}{section}
        \def\theequation{\thesection.\arabic{equation}}}
\numberbysection

\renewcommand{\theequation}{\thesection.\arabic{equation}}
\newcommand{\l@qq}[2]{\addvspace{2em}
 \hbox to\textwidth{\hspace{1em}\bf #1 \dotfill #2}}

\newcounter{app}

\def\app{\setcounter{equation}{0}
\def\theequation{\Alph{app}.\arabic{equation}}\par
   \addvspace{4ex}
   \@afterindentfalse
  \secdef\@app\@dapp}
\newcommand\@app{\@startsection {app}{1}{0ex}%
                                   {-3.5ex \@plus -1ex \@minus -.2ex}%
                                   {2.3ex \@plus.2ex}%
                                   {\normalfont\Large\bf}}

\def\@dapp#1{%
{\parindent \z@ \raggedright  \bf #1}\par\nobreak}
\def\l@app#1#2{\ifnum \c@tocdepth >\z@
    \addpenalty\@secpenalty
    \addvspace{1.0em \@plus\p@}%
    \setlength\@tempdima{2.5em}%
    \begingroup
      \parindent \z@ \rightskip \@pnumwidth
      \parfillskip -\@pnumwidth
      \leavevmode \bfseries
      \advance\leftskip\@tempdima
      \hskip -\leftskip
      #1\nobreak\hfil \nobreak\hb@xt@\@pnumwidth{\hss #2}\par
    \endgroup\fi}
\newcounter{sapp}[app]

\def\sapp{\def\theequation{\Alph{app}.\arabic{equation}}\par
   \@afterindentfalse
  \secdef\@sapp\@dsapp}
\newcommand\@sapp{\@startsection{sapp}{2}{\z@}%
                                     {-3.25ex\@plus -1ex \@minus -.2ex}%
                                     {1.5ex \@plus .2ex}%
                                     {\normalfont\large\bfseries}}

\def\@dsapp#1{%
{\parindent \z@ \raggedright  \bf #1}\par\nobreak}
\newcommand{\l@sapp}{\@dottedtocline{2}{1.5em}{3em}}

\font\teneuf=eufm10  scaled  1440 
\font\seveneuf=eufm7 scaled  1\@ptsize00 
\font\fiveeuf=eufm5  scaled  1\@ptsize00 

\newfam\euffam
\textfont\euffam=\teneuf  \scriptfont\euffam=\seveneuf
  \scriptscriptfont\euffam=\fiveeuf

\def\hexnumber@#1{\ifnum#1<10 \number#1\else
 \ifnum#1=10 A\else\ifnum#1=11 B\else\ifnum#1=12 C\else
 \ifnum#1=13 D\else\ifnum#1=14 E\else\ifnum#1=15 F\fi\fi\fi\fi\fi\fi\fi}

\def\got{\ifmmode\let\next\got@\else
 \def\next{\errmessage{Use \string\got\space only in math mode}}\fi\next}
\def\got@#1{{\got@@{#1}}}
\def\got@@#1{\fam\euffam#1}

\newfont{\lgot}{eufm10 scaled 1440}%

\font\sevenmsa=msam6 
\newfam\msafam
\textfont\msafam=\sevenmsa
\def\hexnumber@#1{\ifnum#1<10 \number#1\else
\ifnum#1=10 A\else\ifnum#1=11 B\else\ifnum#1=12 C\else
\ifnum#1=13 D\else\ifnum#1=14 E\else\ifnum#1=15 F\fi\fi\fi\fi\fi\fi\fi}
\def\msa@{\hexnumber@\msafam}
\def\llcorner{\delimiter"4\msa@78\msa@78 }
\def\lrcorner{\delimiter"5\msa@79\msa@79 }
\mathchardef\blacktriangleright="3\msa@49
\mathchardef\blacktriangleleft="3\msa@4A
\mathchardef\trianglerighteq="3\msa@44
\mathchardef\trianglelefteq="3\msa@45
\font\tenmsb=msbm10 scaled 1\@ptsize00
\newfam\msbfam
\textfont\msbfam=\tenmsb
\def\msb@{\hexnumber@\msbfam}
\mathchardef\varkappa="0\msb@7B

\def\N2{${\cal N}=2$}

\def\theequation{\thesection.\arabic{equation}}

\begin{document}

\thispagestyle{empty}

\begin{flushright}
ITEP-TH-18/00\\
hep-th/0005138
\end{flushright}

\vspace{2.0cm}

\centerline{\Huge Comments on the Morita Equivalence}

\vspace{0.5cm}

\bigskip

\setcounter{footnote}{0}

\setcounter{equation}{0}

\bigskip

\begin{center}

Kirill Saraikin\footnote{e-mail: saraikin@itp.ac.ru}\\

\bigskip

\bigskip

{\em L.D.Landau Institute for Theoretical Physics, 117334, Moscow,
Russia\\

and\\

Institute of Theoretical and Experimental Physics, 117259, Moscow,
Russia}

\end{center}

\bigskip

\bigskip

\setcounter{equation}{0}

\begin{quotation}
\noindent
It is known that noncommutative Yang-Mills  theory
with periodical boundary conditions on torus
at the rational value of the noncommutativity parameter
is Morita equivalent to the ordinary Yang-Mills theory
with twisted boundary conditions on dual torus.
We present simple derivation of this fact.
We describe one-to-one correspondence between
and gauge invariant observables in these two theories.
In particular, we show that under Morita map Polyakov
loops in the ordinary YM theory go to the open
noncommutative Wilson loops discovered by Ishibashi,
Iso, Kawai and Kutazawa.

\end{quotation}

\newpage

\section{Introduction}

Noncommutative geometry deals with functions on deformation
of ordinary space, such that coordinates on it do not
commute\footnote{In the sequel we use the same notation $[~,~]$ both
for ordinary and for star-commutator. To avoid confusions, we supply all
noncommutative quantities with the hats.}:

\be
[\h x_\m, \h x_\n]=2 \pi i \t_{\m\n}, \quad \m,\n=1, \dots d
\ee
The antisymmetrical tensor $\t_{\m\n}$ is called noncommutativity
parameter. Such deformed flat ($\t_{\m\n}=\const$) and compact space
is called noncommutative (quantum) torus ${\bf T}^d_{\t}$.  In
the last few years noncommutative geometry, and especially the noncommutative
torus has been realized to play an important role in compactifications of
M-theory \cite{CDS} and in string theory (see \cite{SW} and references
therein).  It also turned to be very useful in compactification of
instanton's moduli spaces \cite{NS}. The way to deal with the curved quantum
spaces is provided by the Kontsevich's deformation quantization.

A very intriguing subject from noncommutative geometry is
so-called {\it Morita equivalence}~\cite{Sch}.  Roughly
speaking, it states that certain bundles on different noncommutative tori are
dual to each other.  From the physical point of view it results
in equivalence between certain noncommutative and ordinary gauge theories.
In what follows we try to clarify this statement using a
set of simple examples.

\section{Notations}

The algebra $\mathcal A_\t$ of smooth functions on the noncommutative
torus is defined using the Moyal star product:

\be
f \star g \; ( {\bf \h x})=\left.
e^{i \pi \t_{\m\n} \frac{\d}{\d \xi_\m}
\frac{\d}{\d \eta_\n} } f({\bf \xi}) g({\bf \eta})
\right|_{\bf \xi=\eta=\h x}
\ee
The main property of this product is associativity.
In applications it is useful to decompose functions on noncommutative
torus into the Fourier components\footnote{Without loss of generality
we can consider a torus of size $2 \pi$. }:

\be
\label{f}
f({\bf \h x})=\sum_{{\bf k} \in \Z^d} f_{\bf k} e^{i \bf k\h x}
\ee
This corresponds to the Weil or symmetric ordering of coordinates.
Exponents $\h U_{\bf k}= e^{i \bf k\h x}$ may serve as a
basis elements for the algebra $\mathcal A_\t$.

A very intriguing thing happens when
components of the $\t-$tensor becomes rational. Let us first consider
two-torus ${\bf T}^2$:

\be
[\h x_{\m}, \h x_{\n}]=2 \pi i \t \ep_{\m\n}, \quad \m,\n=1,2
\ee
with the rational noncommutativity parameter $\t=\frac{M}{N}$,
where $M$ and $N$ are relatively prime integers.
Then
\be
\label{vrtxalg}
[\h U_{\bf n},\h U_{\bf n'}]=
2i \sin \left( \pi M \frac{n_2n'_1-n_1n'_2}{N} \right) \h U_{\bf n+n'}
= 2i \sin({\bf n} \times {\bf n'}) \, \h U_{\bf n+n'}

\ee
where by definition,
$ {\bf n} \times {\bf n'} \equiv -\pi \t_{\m\n} n_\m n'_\n $.
Note that elements $\h U_{N\bf k}$ generate a center of the $\A_\t$, that
is for any $f({\bf \h x})$:

\be
\left[ e^{i N \bf k \h x },f({\bf \h x})\right]=0
\ee
This means that one can treat  exponents
$\{ \h U_{\bf k}, \ {\bf k}=0|_{\mod N} \}$  in the decomposition (\ref{f})
as if they are ordinary exponents defined on ordinary (commutative) space.
Other $N^2-1$ exponents, obtained from the set
$\{ \h U_{\bf k}, \ {\bf k}\not=0|_{\mod N}\}$ after factorization
over commutative part, generates
closed algebra under star-commutator.
This algebra is isomorphic to the algebra  of $SU(N)$, as we will
see in a moment.
Therefore, at the rational value of the noncommutativity parameter
one can identify algebra of functions on the
noncommutative torus with the algebra of matrix-valued functions
on commutative torus.

We conclude this section by giving an explicit matrix representation
for the noncommutative exponents algebra (see also \cite{Landi}). Such a
representation has been indeed well-known for many
years \cite{PQ,Z}. Let us introduce the following clock and shift generators

\be Q=\left({\begin{array}{lllll} 1& & & & \\
&\omega& & & \\ & &\omega^2& & \\ & & &\ddots& \\ & & & &\omega^{N-1}
\end{array}} \right) ~~~~~~~~~~~~
P=\left({\begin{array}{llllll} 0&1& & & &0\\
&0&1& & & \\
& &\ddots&\ddots& & \\
& & &\ddots& & 1\\
1& & & & &0
\end{array}}\right) \
\ee
where  $\o=e^{2 \pi i\t}$. Matrices $P$ and $Q$ are unitary, traceless
and satisfy:

\be
P^N=Q^N={\bf 1}, \quad PQ=\o~QP
\ee
Moreover,

\be
\label{tr}
\Tr (P^n Q^m)=\left\{ \begin{array}{l} N,\quad if \quad
 n=0|_{\mod N} \quad and \quad m=0|_{ \mod N}  \\
0, \quad if \quad  n\not =0|_{ \mod N} \quad or \quad  m \not =
0|_{ \mod N}
\end{array} \right.
\ee
It is straightforward to check that the generators, defined as

\be
\label{Jn}
J_{\bf n}=\o^{\frac{n_1n_2}{2}} Q^{n_1} P^{n_2}, \quad
{\bf n}=(n_1,n_2)
\ee
satisfy commutation relations (\ref{vrtxalg}):

\be
[J_{\bf n},J_{\bf n'}]=2i \sin \left({\bf n \times n'} \right)
J_{\bf n+n'}
\ee
This identity can be tautologically rewritten in the form of the
Lie algebra commutation relations:

\be
[J_{\bf n},J_{\bf m}]=f_{\bf nm}^{\bf k} J_{\bf k},
\ee
where the structure constants $f_{\bf nm}^{\bf k}$ are

\be
f_{\bf nm}^{\bf k} =2i \dd_{\bf n+m,k} \sin (\bf n \times m)
\ee
The set of unitary unimodular $N \times N $ matrices (\ref{Jn})
suffices to span the algebra of $SU(N)$.

\section{Morita Equivalence}

\subsection{Two-torus. $U(1)|_{\t=\frac{M}{N}} \to U(N) $ }

To define Morita map  we make an additional decomposition
of the function (\ref{f}) on the noncommutative two-torus:

\be
\label{fncomm}
\h f=\sum_{k \in \Z^2} e^{iN \bf k \h x}
\sum_{n_1,n_2=0}^{N-1} f_{\bf k,n} e^{i n_1\h x_1+in_2\h x_2}
\ee
Then we define corresponding $U(N)$-valued function on
the ordinary two-torus as follows:

\be
\label{fcomm}
f=\sum_{k \in \Z^2} e^{iN \bf k  x}
\sum_{n_1,n_2=0}^{N-1} f_{\bf k,n}\ e^{i \bf nx} \ J_{\bf n}
\ee
Because of the relation

\be
\label{mprod}
J_{\bf n} J_{\bf n'} = e^{i \bf n \times n'} J_{\bf n+ n'}
\ee
Morita map
(\ref{fncomm}, \ref{fcomm}) takes star-product to the matrix product.
Obviously, $U(N)$-valued function of general type can not be represented in
the form (\ref{fcomm}). It turns out that this particular form corresponds
to the functions with nontrivial boundary conditions.
Namely, under shifts these functions transforms as

\be
\label{bc}
f(x_1+2\pi \frac{ M}{N},x_2)=\O_1 \ f(x_1,x_2) \ \O_1^{\dagger}, \quad
f(x_1,x_2+ 2\pi \frac{M}{N})=\O_2\ f(x_1,x_2) \ \O_2^{\dagger}
\ee
where

\be
\label{twistmat}
\O_1=(P)^M, \quad \O_2=(Q^{\dagger })^M
\ee
This can
be treated as a constant gauge transformation.
The size $2\pi \frac{M}{N}$ of the dual torus can be
fixed by the requirement for the Morita map to be
single-valued\footnote{I am
indebted to K.  Selivanov for this comment.}.
To illustrate this, let us consider a torus of the size $2\pi \frac{M}{N} n$
(where $n \in {\bf N}$; there are no other possibilities if we want
functions of the type (\ref{fcomm}) to be gauge transformed by
the constant matrix when 
$x$ is shifted by a period of the torus.)
Obviously, in this case there are functions which cannot be represented in
the form (\ref{fcomm}). Such functions do not conjugates when translated
along the vectors $(2\pi \frac{M}{N},0)$ and $(0,2\pi \frac{M}{N})$.

Therefore,
having a set of Fourier coefficients $f_{\bf k,n}$, we
can construct both a function on the noncommutative torus of size $l$
and a matrix-valued function with twisted boundary conditions
(\ref{bc}) on the commutative torus of size $\frac{M}{N} l$ by the
following rule:

\beq
\label{map1}
\left\{
\bal
e^{i{\bf n \h x}} \ \leftrightarrow \  e^{i{\bf n x}} \, J_{\bf n}, \quad
n_1,n_2 < N \\ \\
e^{iN {\bf k \h x}} \ \leftrightarrow  \ e^{i N {\bf k x}} \; {\bf 1}
\eal
\right.
\eeq

\subsection{${\bf T}^d$. $U(1)|_{\t} \to
U(N_1)\times \dots \times U(N_r)$.}

Generalization to the $d$-dimensional case goes by simple
modifications in formulas from the previous subsection.
It is always possible to rotate $\t_{\m\n}$ into a canonical
skew-diagonal form:

\be
\t_{\m\n}=\left(
\begin{array}{lllllll}
~~0 & \t_1 & & & & &\\
-\t_1 & 0 & & & & &\\
& & & \ddots & & &\\
& & & & ~~0 & \t_r &  \\
& & & & -\t_r & 0 & \\
& & & & & & {\bf 0}_{d-2r}
\end{array}
\right)
\ee
where $r$ is the rank of $\t_{\m\n}$. Thus, algebra of higher dimensional
noncommutative torus becomes embedded into a $d$-fold tensor product
of $r$ noncommutative two-tori algebras and ordinary $(d-2r)$-torus
commutative algebra. This immediately leads to other examples of
Morita equivalence, when some of these noncommutative two-tori
are mapped to the commutative ones using relations
(\ref{map1}). If $\t_i=\frac{M_i}{N_i}$, after
Morita map we obtain an ordinary YM theory with the
gauge group $U(N_1)\times \dots \times U(N_r)$.

\subsection{${\bf T}^d$. $U(1)|_{\t} \to U(N) $.}

Algebra of noncommutative
exponents can also be realized using a set of $SU(N)-$valued matrices
$\O_\m, \ \m=1, \dots ,d\quad$  obeying the following relations:

\be
\O_\m\O_\n =e^{2\pi i \t_{\m\n}} \O_\n\O_\m
\ee
Explicit construction of such matrices  can be found
in \cite{Twisteat}. Define generators
$J_{\bf n}$ as follows:

\be
J_{\bf n}=\exp \left( \sum \limits_{\n < \m} \t_{\n\m} n_\n n_\m \right) \
\O_1^{n_1} \dots \O_d^{n_d}
\ee
Then

\be
[J_{\bf n},J_{\bf m}]=2i \sin\left( {\bf n \times m }\right) J_{\bf n+m}
\ee
which coincides with the algebra of noncommutative exponents.
Therefore, in this case Morita map takes the form:

\be
\h f=\sum_{k \in \Z^d} e^{iN \bf k \h x}
\sum_{{\bf n }<N^{\otimes d}} f_{\bf k,n} e^{i \bf n\h x}
\quad \leftrightarrow \quad
f=\sum_{k \in \Z^d} e^{iN \bf k  x}
\sum_{{\bf n }<N^{\otimes d}} f_{\bf k,n}\ e^{i \bf nx} \ J_{\bf n}
\ee

\section{Noncommutative YM vs Ordinary YM}

Let us now turn to the physical applications of the Morita map.
One can define noncommutative version of the Yang-Mills theory
with the action

\be
S_{YM}=\frac{1}{4\pi g^2_{YM}} \int d {\bf x} \; \Tr ( F_{\m\n} F^{\m\n})
\ee
just by replacing in all formulas matrix product by
the Moyal star-product and supplementing all quantities with the hats.
Therefore, noncommutative $U(1)$ Yang-Mills action is

\be
\hat S=\frac{1}{4 \pi g^2_{NC YM}} \int d {\bf \h x} \; \hat F_{\m\n} \star
\hat F^{\m\n}
\ee
where $\h F_{\m\n}=\d_\m \h A_\n -\d_\n \h A_\m -
i [\h A_\m,\h A_\n]_{\s}$.
For simplicity in this section we consider only two-torus.
Generalization to the higher-dimensional case is straightforward.

Morita map takes  NC $U(1)$ gauge fields to the $U(N)$
gauge fields with nontrivial boundary conditions.
Generally, functions on torus can became gauge 
transformed when shifted by a period of the torus:

\be
A_\l({\bf x+l_\m})=\O_\m({\bf x}) \, A_\l({\bf x})\, \O_\m^{-1}({\bf x})  +
i\O_\m({\bf x})\, \d_{\l}\, \O_\m^{-1}({\bf x})
\ee
where $\O_\m({\bf x})$ are elements of the $U(N)$ group, known
as twist matrices. They should satisfy a consistency conditions:

\be
\label{cnds}
\O_\m({\bf x+l_\n})\, \O_\n({\bf x})=e^{2\pi i \frac{M}{N}\ep_{\m\n}} \,
\O_\n({\bf x+l_\m})\, \O_\m({\bf x})
\ee
An integer $M$ in this formula is so-called 't Hooft's flux.
It is known only three types of possible boundary conditions
(solutions of the eqs (\ref{cnds}) ):

\noindent
1. {\it twist eaters}: $\O_\m=\const$ \\
2. {\it abelian twists} \\
3. {\it nonabelian twists}\\
For more details see the recent review \cite{GA}.

The map (\ref{map1}) corresponds exactly to the
first case.
It is not well understood how to realize Morita map corresponding
to the other boundary conditions.
Roughly speaking, when working in the
Fourier basis (\ref{f}), after shifts one can only multiply functions on
numbers and cannot add something like $\O_\m({\bf x})\, \d_{\l}\,
\O_\m^{-1}({\bf x})$. To do this, one needs another basis for the functions
on noncommutative torus (creation/annihilation operators, noncommutative
theta-functions?).

Under Morita map, defined in the previous section,
actions go to the actions,
equations  of motions go to the equations of motions, and
solutions (e.g. instantons) also go to the solutions, even
at the quantum level.
These properties of the Morita map can be encoded in the
following identity:

\be
\label{idt}
\int d{\bf \h x} \ \h A_\m \s \h A_\n \s \dots \s \h A_{\l} =
\frac{1}{N} \int d{\bf x} \ \Tr ( A_\m  A_\n  \dots  A_{\l})
\ee
which is straightforward to prove using the definition

\be
\int d{\bf \h x} \ e^{i\bf k \h x} =\dd_{\bf k,0}
\ee
and the  property (\ref{tr}) of the clock and shift generators.
In fact, one can insert arbitrary number of derivatives
into the integrals in (\ref{idt}) and thus obtain equivalent
gauge invariant quantities in noncommutative and ordinary gauge theories.
Due to the identity (\ref{idt}) we can establish the following
correspondence between correlators:

\be
\label{cor}
\int \D A^\m_{\bf k,n}\ e^{\h S[\t=\frac{M}{N}]}
\ \h \OO_1 \dots \h \OO_l
 = \int \left. \D A^\m_{\bf k,n} \ e^{ S_{YM}}
\right|_{fxd~bndry~conds,~flux=M} \ \OO_1 \dots \OO_l
\ee
where $g^2_{NC YM}=N g^2_{YM}$, and

\be
\h \OO =\int d \h {\bf x} \  (\h F_{\m\n})^{\star n}, \quad
\OO=\frac{1}{N} \int d {\bf x} \, \Tr \, (F_{\m\n})^{ n}
\ee
Other important gauge invariant
quantities of the YM theory are the Wilson
loops:

\be
W[C]=\Tr \, P\exp \left( i \, \oint_C A_\m({\bf x}) \, dx_\m \right)
\ee
which corresponds to the closed path C. On torus  there are
paths from the different homotopy classes, which can be
classified by winding numbers $w_\m$ around the $\m$-th
direction. The corresponding Wilson loops are called Polyakov
loops. The simplest Polyakov loop corresponds to
the straight line along the   $\m$-th direction:

\be
W_P[{\bf x},\m]=\Tr \left[
P\exp \left(i \, \int \limits_{\bf x}^{\bf x+ l_\m}
A_\m({\bf x}) \, dx_\m \  \right)  \O_\m e^{i x_\m}\right]
\ee
where insertion of the twist matrix (\ref{twistmat})
is necessary to guarantee gauge invariance.

Wilson lines in noncommutative Yang-Mills theory were constructed by
Ishibashi, Iso, Kawai and Kutazawa \cite{IIKK} (see also
\cite{AMNS,AMNSnew}).
This construction goes as follows. First, introduce an oriented curve
$C$ in auxiliary commutative two-dimensional space parametrized
by the functions $\xi(\sigma)$ with $0\le \sigma \le 1$. Fix the starting
point $\xi_\m(0)=0$ end the endpoint $\xi_\m(1)=v_\m$.
Then, assign to this curve a
noncommutative analog of the parallel transport operator:

\be
\label{Wn}
{\mathcal U}[{\bf \h x}, C]= 1+
\sum_{n=1}^{\infty} i^n \int \limits_{0}^{1} d\sigma_1
\int \limits_{\sigma_1}^{1} d\sigma_2
\ \dots
\int \limits_{\sigma_{n-1}}^{1} d\sigma_n \;
\frac{d \xi_{\m_1}(\sigma_1)}{d \sigma_1} \dots
\frac{d \xi_{\m_n}(\sigma_n)}{d \sigma_n} \\
\; \times A_{\m_1}({\bf \h x+\xi(\sigma_1) })  \s
\dots \s A_{\m_n}({\bf \h x+\xi(\sigma_n) })
\ee
The series in (\ref{Wn}) is noncommutative analog of the
$P$-exponent. The star-gauge invariant quantity is then

\be
\h \OO [C] =\int d{\bf \h x} \
{\mathcal U}[{\bf \h x}, C] \s S[{\bf \h x}, C]
\ee
where  $S[{\bf \h x}, C]=1$ if the path $C$ is closed and

\be
S[{\bf \h x}, C] =e^{i(\t^{-1})_{\m\n} v_\n \h x_\m}
\ee
if the path is open.
Gauge invariance requires that the coordinates of the endpoint must be equal
to $v_\m=2\pi r_\m \frac{M}{N}, \quad r_\m=0, \dots, N-1$.
In the simplest case, when $C_\m$ is the straight line
along the   $\m$-th direction and $v_\m=2\pi \frac{M}{N}$,
the function $S[{\bf \h x}, C_\m]$
under Morita map (\ref{map1})   go to the
twist function $\O_\m e^{i x_\m}$. Therefore, using identity (\ref{idt}) we
obtain the following relation between the Polyakov loops in the ordinary YM
theory and open noncommutative Wilson loops:

\be
\frac{1}{N}\int   d{\bf x} \ W_P[{\bf x},\m]=\h \OO [C_\m]
\ee

\bigskip

\section{Conclusions}

In this paper we have made some comments on the Morita equivalence
between noncommutative and ordinary gauge theories.
We present a simple prescription how to identify gauge
fields and correlators of the gauge invariant observables in
the $U(1)$ NC YM theory on torus at the rational value of the
$\t$-parameter
with those ones in the ordinary
$U(N)$ or $U(N_1) \times \dots \times U(N_r)$ YM theory with nontrivial
boundary conditioxns on the dual torus.
The size of the dual torus is determined by the requirement for
the Morita map to be single-valued.
We also show that under Morita map Polyakov loops in the
ordinary YM theory go to the open noncommutative Wilson
loops\footnote{This fact firstly was mentioned in \cite{AMNSnew}}.

An open question is to generalize Morita equivalence to the case of
the non-twist-eater's type boundary conditions.
Another interesting direction is to link three different
descriptions of the Morita equivalence:
field theory approach using the Fourier components, string
theory approach using T-duality and brane language \cite{PSch,Ohta},
and mathematical approach via the twisted bundles over
the noncommutative torus \cite{Sch,HV}.

\bigskip

\bigskip

\centerline{\large \bf Acknowledgements}

\bigskip

I am grateful to D. Belov and N. Nekrasov for important comments
reviving the interest in the subject.
I thank S. Gorsky for numerous discussions and correspondence,
and A. Morozov for his interest in this work and encouragement.
I am also grateful to I. Polyubin, A. Rosly and especially to K. Selivanov for
useful comments on the manuscript. I acknowledge Y. Makeenko for helpful
discussion. I would like to thank Ira Vashkevich for technical support.
The work was partly supported by the Russian President's grant
00-15-99296 and RFBR grant 98-02-16575.

\end{document}